\documentclass[aps,reprint,groupedaddress]{revtex4-1}
\usepackage{amsmath}
\usepackage{bm} 
\usepackage{amssymb}
\usepackage{color}
\usepackage{multirow}
\usepackage[colorlinks,bookmarks=false,citecolor=red,linkcolor=blue,urlcolor=blue]{hyperref}
\usepackage{graphicx}
\usepackage{mathrsfs,bbm}
\usepackage[caption=false]{subfig} 
\usepackage{wrapfig}
\usepackage[version=3]{mhchem}
\usepackage{dcolumn}
\usepackage{calligra}
\DeclareMathAlphabet{\mathcalligra}{T1}{calligra}{m}{n}
\DeclareFontShape{T1}{calligra}{m}{n}{<->s*[2.2]callig15}{}

\def\A{{\bf A}}

\def\S{{\bf S}}

\def\k{{\bf k}}
\def\r{{\bf r}}

\def\dr{{\bf dr}}
\def\bmdelta{\bm{\delta}}
\def\bmsigma{\bm{\sigma}}
\def\bmtau{\bm{\tau}}
\def\bmdelta{{\bm \delta}}

\def\sbar{\bar{s}}
\def\sfz{\mathsf{z}}
\def\Hhat{\hat{H}}

\def\ahat{\hat{a}}
\def\bhat{\hat{b}}
\def\chat{\hat{c}}

\def\nhat{\hat{n}}

\def\that{\hat{t}}
\def\xhat{\hat{x}}

\def\zhat{\hat{z}}
\def\etahat{\hat{\eta}}

\def\phihat{\hat{\phi}}
\def\psihat{\hat{\psi}}

\def\nbar{\bar{n}}

\begin{document}
\title{Theory of quantum oscillations of magnetization in Kondo insulators}
\author{Panch Ram} 
\author{Brijesh Kumar}
\email{bkumar@mail.jnu.ac.in}
\affiliation{School of Physical Sciences, Jawaharlal Nehru University, New Delhi 110067, India.}
\date{\today} 


\begin{abstract}
The Kondo lattice model of spin-1/2 local moments coupled to the conduction electrons at half-filling is studied for its orbital response to magnetic field on bipartite lattices. Through an effective charge dynamics, in a canonical representation of electrons that appropriately describes the Kondo insulating ground state, the magnetization is found to show de Haas-van Alphen oscillations from intermediate to weak Kondo coupling. These oscillations are ascribed to  the inversion of a dispersion of the gapped  charge quasiparticles, whose chemical potential surface is measured by the oscillation frequency. Such oscillations are also predicted to occur in spin-density wave insulators. 
\end{abstract}

\pacs{75.10.Jm, 75.10.Kt, 75.30.Kz, 05.30.Rt}

\maketitle


{\em Introduction.---} 
Typically realised in rare-earth compounds, the Kondo insulators are dense arrays of local moments interacting with the conduction electrons at half-filling~\cite{Coleman.Book.2015,Misra.Book.2008,Aeppli.Fisk.1992}. They exhibit insulating behavior at low temperatures due to singlet formation between the local moments and the conduction electrons. Recent observations of de Haas-van Alphen 
oscillations in \ce{SmB6} has greatly renewed the interest in Kondo insulators~\cite{Li2014,Tan2015}. 

The de Haas-van Alphen (dHvA) effect refers to the quantum 
oscillation of magnetization as a function of the (inverse) magnetic field. 
It is considered a hallmark of the metallic response, and a direct probe of the Fermi surface (FS)~\cite{Ashcroft.Mermin, Abrikosov1988,Onsager}. 
The dHvA oscillations are a manifestation of the Landau quantization of electronic states in uniform magnetic field. An insulator is not expected to show dHvA oscillations. 
But the case of \ce{SmB6} presents a counterexample to this conventional view, and poses a question of principle on the occurrence of dHvA oscillations in the insulators. This question has been given some attention recently, with some studies getting the hitherto unexpected dHvA oscillations in mostly the band-theoretic models of insulators~\cite{Kishigi2014,Knolle2015, Zhang2016, Baskaran2015, Erten2016, Pal2016}. But the situation in a Kondo insulator (KI) is more precarious, where the electrons are correlated and localized, and one is not quite sure which quasiparticles, if any, cause dHvA oscillations, and what surface, Fermi or otherwise, is being measured. 
 
Topologically protected conducting surface states in a topological Kondo insulator with an insulating bulk could in principle give quantum oscillations~\cite{Dzero2016, Erten2016}. 
 But the FS measured from quantum oscillations in \ce{SmB6} corresponds to the half of its bulk Brillouin zone (BZ)~\cite{Tan2015}. This can not be accounted for by the surface states, and
calls for an understanding of the dHvA oscillations within the bulk insulating behaviour of the KI's. 

 Another scenario treats the Kondo insulating state on bipartite lattice (\ce{SmB_6} has a simple cubic structure) at half-filling as a scalar Majorana Fermi sea spread over half of the bulk BZ~\cite{Baskaran2015}. While it may look agreeable on the size of the observed FS, it has gapless quasiparticles, and this gapless Majorana sea can not describe an insulator~\footnote{The Fermi sea of non-interacting electrons on bipartite lattice, which is a conducting state, consists of four such independent gapless Majorana Fermi seas.}. A recent experiment {\color{red} also} rules this out~\cite{Xu2016}. 
  
In this paper, we  study the Kondo lattice model using a canonical representation of electrons~\cite{BK2008} that appropriately describes the Kondo insulating state on bipartite lattices, and gives the 
 quantum oscillations of magnetization as a general bulk property for the Kondo coupling ranging from intermediate to weak. We  get these oscillations due to 
the inversion of a dispersion  of the charge quasiparticles whose effective chemical-potential surface they measure.
These quasiparticles 
are gapped and occupy half of the bulk BZ. This approach also applies to the Hubbard model, 
and predicts the dHvA oscillations to occur in the insulating spin-density wave state. 
 

{\em Kondo Lattice Model.---} 
To understand dHvA oscillations in Kondo insulators, we study the orbital response to magnetic field in the ground state of the basic Kondo lattice model (KLM), $\Hhat$, of local spin-1/2 moments coupled via antiferromagnetic exchange, $J>0$, to the conduction electrons at half-filling with nearest-neighbour hopping, $t>0$, on square and simple cubic lattices.
\begin{align}
 \Hhat &=-t\sum_{\r,\bmdelta,s} e^{i\frac{e}{\hbar}\int_\r^{\r+\bmdelta}\A\cdot \dr}\chat^\dag_{\r,s} \chat^{ }_{\r+\bmdelta,s} + \frac{J}{2} \sum_\r \S_\r\cdot\bmtau^{ }_\r 
\label{eq:H}
\end{align}
Here, $\r$ runs over the lattice sites, $\bmdelta$ is summed over the nearest neighbours of $\r$, and $s=\uparrow,\downarrow$ is the spin label. 
The $\chat_{\r,s}$ ($\chat^\dag_{\r,s}$), are the annihilation (creation) operators of the conduction electrons, whose spin operators are denoted as $\S_\r$. 
The Pauli operators, $\bmtau_\r =(\tau^x_\r, \tau^y_\r, \tau^z_\r)$, denote the local moments. The uniform external magnetic field, $B\zhat$, is coupled here to the electronic motion via Peierls phase in terms of the vector potential, $\A=-By \xhat$.  

To set up our scheme of calculation, we first discuss the KLM without magnetic field. A canonical representation in terms of spinless fermions and Pauli operators has been found to be fruitful in describing correlated electrons~\cite{BK2008,BK2009,BK2013}. Following Ref.~\onlinecite{BK2008}, we employ it here to rewrite the KLM for $B=0$ on bipartite lattice as follows:  
\begin{eqnarray}
\Hhat &=& -\frac{it}{2} \sum_{\r\in \mathcal{A}} \sum_\bmdelta \left[ \psihat_{a,\r}\phihat_{b,\r+\bmdelta} + \psihat_{b,\r+\bmdelta}\phihat_{a,\r} \left(\bmsigma_\r\cdot\bmsigma_{\r+\bmdelta}\right)\right] \nonumber \\
&& + \frac{J}{4}\left[\sum_{\r\in \mathcal{A}} \nhat_{a,\r} (\bmsigma_\r \cdot \bmtau_\r) + 
\sum_{\r\in \mathcal{B}} \nhat_{b,\r} (\bmsigma_\r\cdot \bmtau_\r)\right],
\label{eq:Hcanon}
\end{eqnarray}
where $\phihat_{a,\r} = \ahat^\dag_\r+\ahat^{ }_\r$ and $i\psihat_{a,\r} = \ahat^\dag_\r-\ahat^{ }_\r$ are the Majorana operators corresponding to the spinless fermion operators, $\ahat_\r$, on $\mathcal{A}$ sublattice, and likewise, $\phihat_{b,\r}$ and $\psihat_{b,\r}$ for $\bhat_\r$ on $\mathcal{B}$ sublattice. Moreover, $\nhat_{a(b),\r} = \ahat^\dag_\r\ahat_\r \, (\bhat^\dag_\r\bhat_\r)$ are their number operators, and $\bmsigma_\r$'s are the Pauli operators.
In this representation, $\chat^\dag_{\r\uparrow} = \phihat_{a,\r}\sigma^+_\r$, $\chat^\dag_{\r\downarrow} = \frac{1}{2}(i\psihat_{a,\r} - \phihat_{a,\r}\sigma^z_\r)$ and $\S_\r=\frac{1}{2}\nhat_{a,\r} \bmsigma_\r$ on $\mathcal{A}$ sublattice, and $\chat^\dag_{\r\uparrow} = i\psihat_{b,\r}\sigma^+_\r$, $\chat^\dag_{\r\downarrow} = \frac{1}{2}(\phihat_{b,\r} - i\psihat_{b,\r}\sigma^z_\r)$ and $\S_\r=\frac{1}{2} \nhat_{b,\r} \bmsigma_\r$ on $\mathcal{B}$ sublattice~\cite{BK2008}. Moreover, the number operator for total $\uparrow$ and $\downarrow$ electrons on a site $\r\in \mathcal{A} (\mathcal{B})$ is: $1+\sigma^z_\r (1-\nhat_{a(b),\r})$. 

The form of Eq.~(\ref{eq:Hcanon}) clearly suggests that, if $J \gg t$, then $\bmsigma_\r$ and $\bmtau_\r$ would locally form singlet in the ground state, while the spinless fermions describe the residual charge dynamics through, $\Hhat_0 = - \frac{it}{2}\sum_{\r\in \mathcal{A}}\sum_{\bmdelta} \psihat_{a,\r}\phihat_{b,\r+\bmdelta} -\frac{3J}{4}(\sum_{\r\in \mathcal{A}} \nhat_{a,\r} + \sum_{\r\in \mathcal{B}} \nhat_{b,\r}) $, which has a charge gap, $\Delta_c = \sqrt{(3J/4)^2+(\sfz t/2)^2}-\mathsf{z}t/2$~\footnote{The term corresponding to $t$ in $\Hhat_0$ is the so-called scalar Majorana Fermi sea of Ref.~\onlinecite{Baskaran2015}. But here it occurs with an additional term due to $J$ that opens the charge gap.}. Here, $\mathsf{z}$ is the nearest neighbour coordination. This singlet state also has a spin gap, $\Delta_s = J$, and keeps the local occupancy at one electron per site. But for $B\neq 0$, we do not get quantum oscillations in this idealized model of strong-coupling KI. Hence, we improve upon it by correcting the local singlets for the exchange interaction caused by hopping, and also correcting in return the charge dynamics self-consistently. 

To this end, we decouple the Pauli operators from the spinless fermions  in Eq.~\eqref{eq:Hcanon} write an approximate version of the KLM: $\Hhat \approx \Hhat_{c} + \Hhat_{s} + e_1 L$, with
\begin{subequations}
\label{eq:HabHst}
\begin{eqnarray}
\Hhat_{c} &=& -\frac{it}{2} \sum_{\r\in \mathcal{A}} \sum_\bmdelta \left[ \psihat_{a,\r}\phihat_{b,\r+\bmdelta} + \rho_1 \psihat_{b,\r+\bmdelta}\phihat_{a,\r}\right] \nonumber \\
&& +\frac{J \rho_0}{4}\left[\sum_{\r\in \mathcal{A}} \nhat_{a,\r} + \sum_{\r\in \mathcal{B}} \nhat_{b,\r}\right], \\
\Hhat_{s} &=& \frac{t\zeta}{4}\sum_{\r,\bmdelta} \bmsigma_\r\cdot\bmsigma_{\r+\bmdelta} +\frac{J \nbar}{4}\sum_\r \bmsigma_\r\cdot\bmtau_\r,
\end{eqnarray}
\end{subequations}
and $e_1 = -(\sfz t\zeta \rho_1 + J \nbar \rho_0)/4$. Here, $L$ is the total number of sites, $\rho_0=\frac{1}{L}\sum_\r\langle \bmsigma_\r\cdot\bmtau_\r\rangle$, $\rho_1 = \frac{1}{\sfz L}\sum_{\r,\bmdelta}\langle \bmsigma_\r\cdot\bmsigma_{\r+\bmdelta}\rangle$, $\nbar=\frac{1}{L}\langle \sum_{\r\in\mathcal{A}} \nhat_{a,\r} + \sum_{\r\in\mathcal{B}}\nhat_{b,\r}\rangle$ is the density of spinless fermions, and $\zeta=\frac{2i}{\sfz L}\sum_{\r\in\mathcal{A}} \sum_\bmdelta\langle \phihat_{a,\r}\psihat_{b,\r+\bmdelta}\rangle$. These mean-field parameters, $\rho_0$, $\rho_1$, $\zeta$ and $\nbar$, are determined self-consistently by solving $\Hhat_{c}$ and $\Hhat_{s}$~\footnote{The $\Hhat_{s}$ resembles the Kondo necklace model~\cite{Doniach-necklace}. Here, it describes the magnetic properties of KI.}. 


{\em Kondo insulator in zero field.---} The  effective charge dynamics of KI in the diagonal form is  given here by \( \Hhat_{c} = J \rho_0 L/8 + \sum_{\k} \sum_{\nu=\pm} E_{\k\nu} (\etahat^\dag_{\k\nu} \etahat^{ }_{\k\nu}-1/2), \)  where $\k \in$ the half-BZ, $E_{\k\pm} = E_\k \pm \frac{1}{2}t(1+\rho_1)|\gamma_k| >0$, $\gamma_\k=\sum_\bmdelta e^{i\k\cdot\bmdelta}$, $E_\k =\sqrt{(J \rho_0/4)^2 + [t(1-\rho_1)|\gamma_\k|/2]^2}$, and $\etahat_{\k\nu}$ are the fermionic quasiparticle operators. The equations for $\nbar$ and $\zeta$ in the ground state of $\Hhat_{c}$ ($i.e.$, the vacuum of the gapped  charged quasiparticles) are: 
\begin{align}
& \nbar = \frac{1}{2} - \frac{J \rho_0}{4L}\sum_\k\frac{1}{E_\k}~ \mbox{and}~ \zeta = \frac{t(1-\rho_1)}{\sfz L}\sum_\k\frac{|\gamma_\k|^2}{E_\k}.
\label{eq:nbar-zeta}
\end{align}

We study $\Hhat_{s}$ using bond-operator representation of $\bmsigma_\r$ and $\bmtau_\r$~\cite{SachdevBhatt,BK2010}, where the effective spin dynamics, $\Hhat_{s} \approx  L[\lambda\sbar^2 -5\lambda/2 - J\nbar(\sbar^2 - 1/4)] + \sum_{\k}\sum_{\alpha=x,y,z} \varepsilon_\k (\that^\dag_{\k\alpha}\that^{ }_{\k\alpha} +1/2)$, is given in terms of the bosonic triplon excitations, $\that_{\k\alpha}$, with respect to the local Kondo singlets of mean amplitude, $\sbar$. Here, $\varepsilon_\k = \sqrt{\lambda(\lambda+t\zeta\sbar^2 \gamma_\k)}$ is the triplon dispersion, with $\lambda$ as Lagrange multiplier and $\k \in$ the full BZ. The mean-field parameters for this part are given as: $\rho_0 = 1-4\sbar^2$ and $\rho_1 = 4\sbar^2(J\nbar - \lambda)/\sfz t\zeta$, where 
\begin{subequations}
\label{eq:sbar-lambda}
\begin{eqnarray}
\sbar^2 &=& \frac{5}{2}-\frac{3}{4L}\sum_\k\frac{2\lambda+t\zeta\sbar^2 \gamma_\k}{\varepsilon_\k},\, \mbox{and} \\
\lambda &=& J\nbar - \frac{3\lambda t \zeta}{4L}\sum_\k \frac{\gamma_\k}{\varepsilon_\k}.
\end{eqnarray}
\end{subequations}

\begin{figure}[t]
\centering
\includegraphics[width=.35\textwidth]{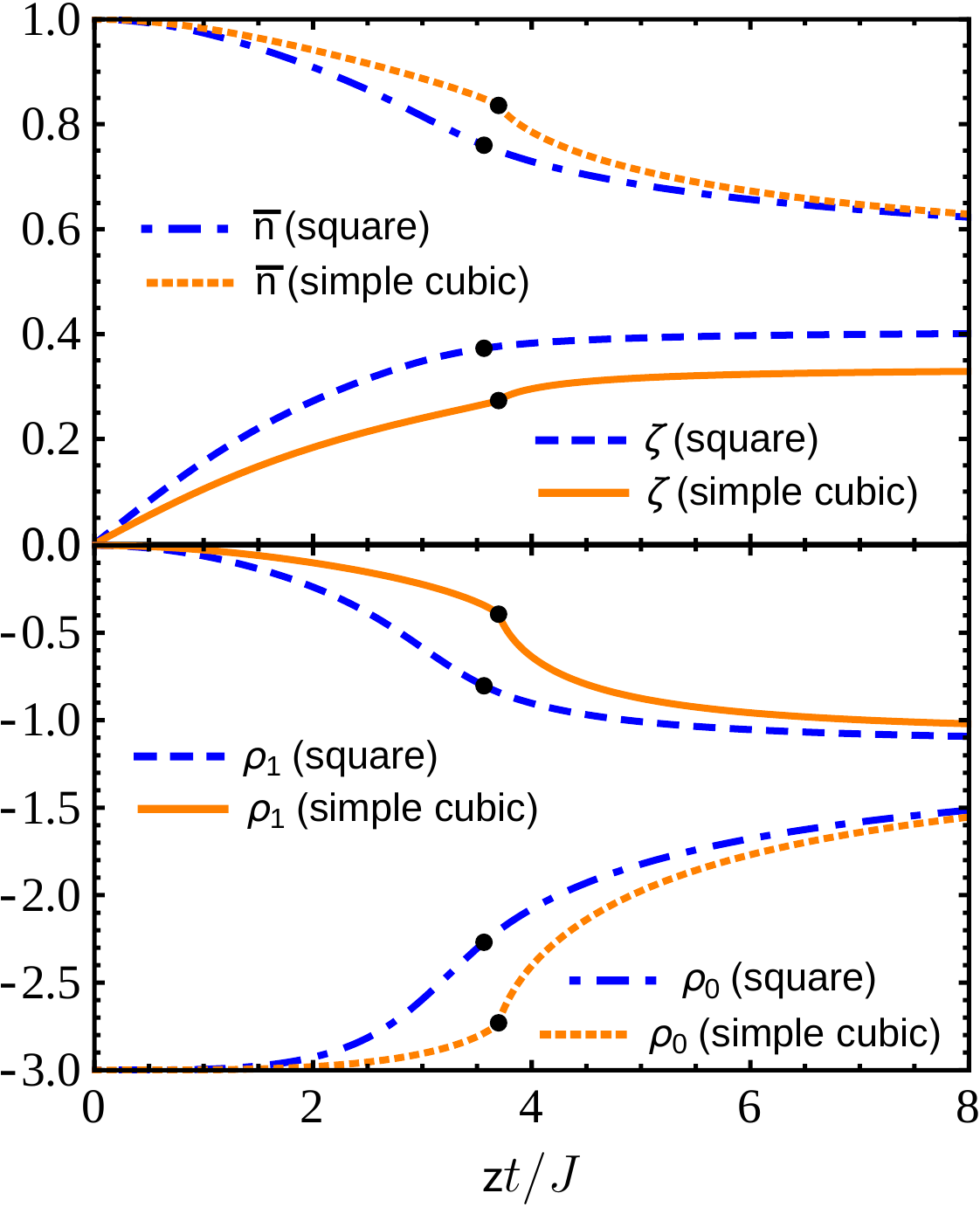}
\caption{Mean-field parameters of the effective charge and spin dynamics as a function of $t/J$ on square and simple cubic lattices. The black dots indicate the critical hopping, $t_c$, below which the insulating ground state is a Kondo singlet, and above which, it is antiferromagnetically ordered [see Fig.~\ref{fig:mf-parameters}(d) for the spin and charge gaps].}
\label{fig0}
\end{figure}

\begin{figure}[htbp]
\centering
\includegraphics[width=.46\textwidth]{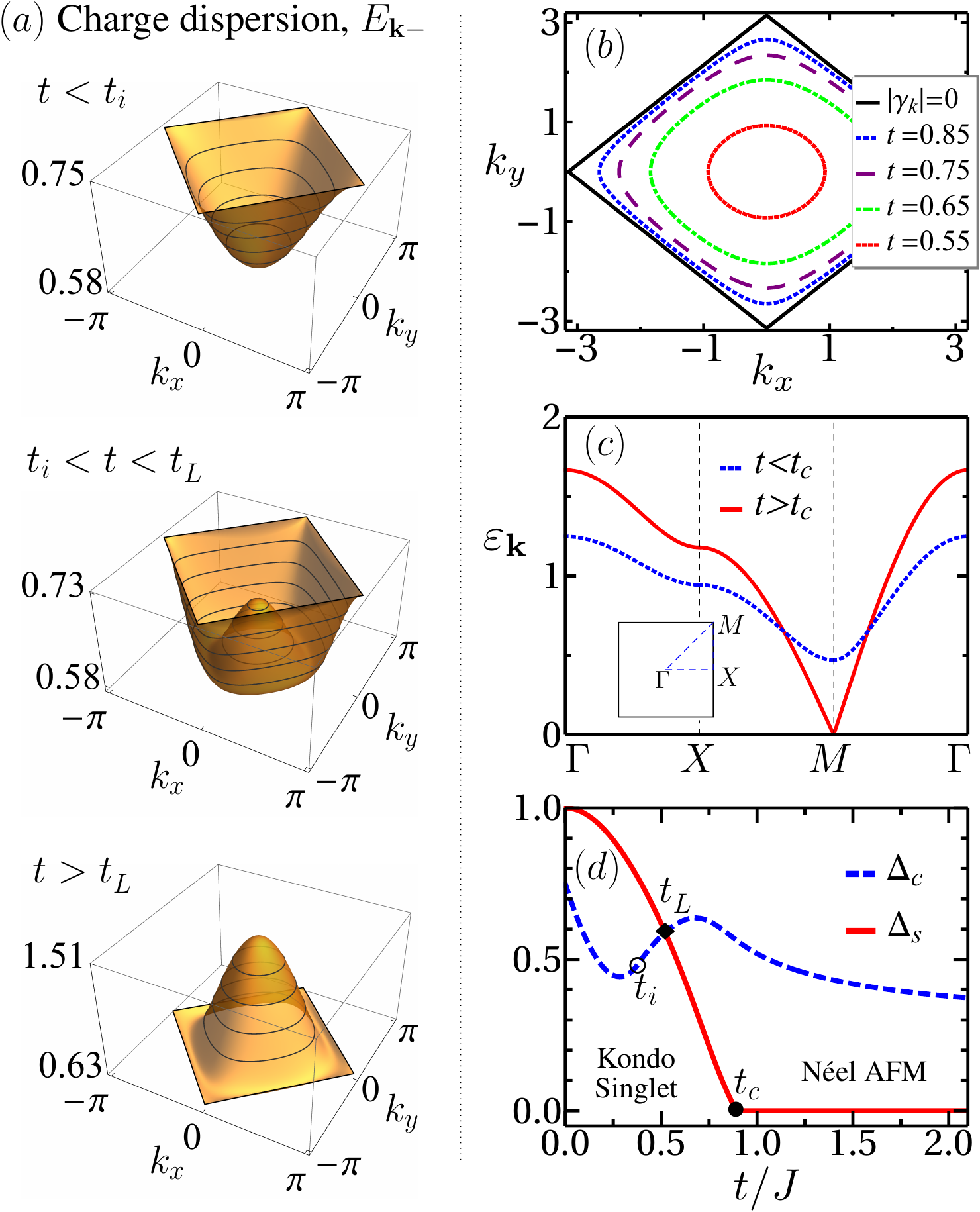}
\caption{Key features of the Kondo insulating ground state from Eq.~(\ref{eq:HabHst}) on square lattice  (with $J=1$). ($a$) Dispersion, $E_{\k-}$, of $\Hhat_{c}$ undergoes \emph{inversion} for $t>t_i$. 
($b$) The \emph{chemical-potential} surface (CPS), $E_{\k-} = J|\rho_0|/4$, for $t > t_L$, where $t_L$ is the point of \emph{Lifshitz}-like transition, below which $|\gamma_\k|=0$ is the CPS, and above $t_L$, the CPS has a second $t$ dependent branch that approaches $|\gamma_\k|=0$ with increasing $t$. 
($c$) Triplon dispersion, $\varepsilon_\k$, of $\Hhat_{s}$. It is gapped (Kondo singlet) for $t < t_c$ and gapless (N\'eel antiferromagnetic) for $t>t_c$. See Eq.~\ref{eq:tstarLc} for $t_i$, $t_L$, and $t_c$. ($d$)  Charge ($\Delta_c$) and spin ($\Delta_s$) gaps vs. $t/J$. 
}
\label{fig:mf-parameters}
\end{figure}

We compute $\nbar$, $\zeta$, $\rho_0$ and $\rho_1$ by solving Eqs.~(\ref{eq:nbar-zeta}) and~(\ref{eq:sbar-lambda}) for different values of $t$, with $J=1$. At $t=0$, their exact values are: $\rho_0=-3$, $\rho_1=0$, $\nbar=1$, and $ \zeta=0$. For $t>0 $, we get $-3 < \rho_0 \lesssim \rho_1 <0 $, and $0 < \zeta < 0.5 < \nbar <1 $, as shown in Fig.~\ref{fig0}. We correctly find the $\Hhat_{c}$ to have a non-vanishing charge gap, $\Delta_c$, whereas $\Hhat_{s}$ exhibits a spin-gap, $\Delta_s$, for $t<t_c$ in the Kondo singlet phase. The $\Delta_s$ goes continuously to zero at $t_c$, causing a transition to N\'eel antiferromagnetic (AFM) state, as shown in Fig.~\ref{fig:mf-parameters}($c$)-($d$) for square lattice. 
Our calculation slightly overestimates the $t_c$, as compared to the values from other methods~\cite{Assaad1999,Shi.Singh.1995,Wang.Li.Lee.1994}.

A special feature of the Kondo insulating state that we discover here is the inversion of a charge quasiparticle dispersion, that has direct bearing on quantum oscillations. The dispersions, $E_{\k \pm} >0$, always touch each other at $|\gamma_\k|=0$, at a value of $J|\rho_0|/4$, which is the chemical potential of the spinless fermions in $\Hhat_{c}$. For small $t/J$, $E_{\k-(+)}$ is lowest (highest) at $\k=0$, and highest (lowest) at $|\gamma_\k|=0$. But for $t > t_i$, the $\k=0$ becomes a point of local maxima of $E_{\k-}$, whose lowest value ($\Delta_c$) now lies on the contour, $|\gamma_\k|=J|\rho_0|(1-|\rho_1|)/\{4t(1+|\rho_1|)\sqrt{|\rho_1|}\}$, while $E_{\k+}$ is always maximum at $\k=0$~\footnote{ No such inversion occurs for the triplon dispersion, $\varepsilon_\k$.}. A similar shift in the band minimum at a similar value of $t_i$ has also been noted in Ref.~\onlinecite{Trebst2006}.  Furthermore, for $t > t_L > t_i$, the $\k=0$ becomes the global maxima of $E_{\k-}$, which leads to a second branch of the \emph{chemical-potential surface} (CPS) given by $|\gamma_\k| = J|\rho_0|(1-|\rho_1|)/\{4t|\rho_1|\}$, in addition to $|\gamma_\k|=0$ [see Figs.~\ref{fig:mf-parameters}($a$-$b$) and~\ref{fig:mag-kondo-cub}($c$-$d$)]. This is akin to Lifshitz transition~\cite{Lifshitz}, but in a Kondo insulator! We will see that, for dHvA oscillations, the CPS in KI plays the role of Fermi surface in metals. 
Sufficiently above $ t_L$, the $E_{\k-}$ nearly fully inverts and looks similar to $E_{\k+}$. This 
inversion of $E_{\k-}$, shown in Fig.~\ref{fig:mf-parameters}($a$) for square lattice, is generic to Kondo insulators, at least on bipartite lattices. Having obtained this novel and other expected features of the KI's using $\Hhat_{c}+\Hhat_{s}$, we now study this minimal approximate model in a uniform magnetic field. 
\begin{equation}
\begin{array}{r|ccc}
&  t_i & t_L & t_c \\ \hline
\mbox{square lattice}~ & 0.38~ & 0.52~ & 0.89  \\ \hline
\mbox{simple cubic lattice}~ & 0.33~ & 0.48~ & 0.62
\end{array}
\label{eq:tstarLc}
\end{equation}
 
{\em The dHvA oscillations in KI.---} 
By rewriting Eq.~(\ref{eq:Hcanon}) for $B\neq 0$, and keeping only those terms that couple to $\nbar$, $\zeta$, $\rho_0$ and $\rho_1$, we get the following $B$ dependent minimal models of charge and spin dynamics of a Kondo insulator.
\begin{subequations}
\label{eq:HabHst-B}
\begin{align}
&\Hhat_{c}^{[B]} = -\frac{it}{2} \sum_{\r\in \mathcal{A}} \sum_\bmdelta \Bigg\{ \left[ \psihat_{a,\r}\phihat_{b,\r+\bmdelta} + \rho_1 \psihat_{b,\r+\bmdelta}\phihat_{a,\r}\right] \times  \nonumber \\
& \cos{\left(2\pi\alpha\, \r_y\, \xhat\cdot\hat{\bmdelta}\right)} \Bigg\} 
+ \frac{J \rho_0}{4}\left[\sum_{\r\in \mathcal{A}} \nhat_{a,\r} + \sum_{\r\in \mathcal{B}} \nhat_{b,\r}\right] \label{eq:Hab-B}\\
& \Hhat_{s}^{[B]} = \frac{t\zeta}{4}\sum_{\r,\bmdelta} \cos{\left(2\pi\alpha\, \r_y\, \xhat\cdot\hat{\bmdelta}\right)} \bmsigma_\r\cdot\bmsigma_{\r+\bmdelta} +\frac{J \nbar}{4}\sum_\r \bmsigma_\r\cdot\bmtau_\r 
\end{align}
\end{subequations}
These are Hofstadter~\cite{Hofstadter1976} type 
 models, but of Majorana fermions and hard-core bosons. Here, $\alpha=e Ba^2/h$ is the magnetic flux, 
$a$ is the  lattice constant, integer $\r_y$ is the $y$-coordinate of $\r$, and $\hat{\bmdelta} = \bmdelta/|\bmdelta|$. We put zero-field values of $\rho_0$, $\rho_1$, $\nbar$ and $\zeta$ in Eqs.~(\ref{eq:HabHst-B}), and compute magnetization, $M = -\partial e_g/\partial \alpha$, as a function of $\alpha=p/q$ for integer $p=1, 2, \dots q$ with $q$ upto $709$ on square lattice, and $401$ on simple cubic lattice~\footnote{ Unlike the basic Hofstadter model, in our $H_c^{[B]}$ (that has hopping and pairing), the Landau bands  are somewhat dispersive with respect to $k_x$, even for large $q$. Hence, in our calculations, we have also taken upto 288 $k_x$-points.}.  
 Here, $e_g$ is the ground state energy per site of Eqs.~\eqref{eq:HabHst-B}. As the contribution to $M$ from $ \Hhat_{s}^{[B]} $ happens to be quite  ($\sim 100$ times) small compared to $\Hhat_{c}^{[B]}$, and we see dHvA oscillations only through charge dynamics, 
below we discuss the results for $\Hhat_{c}^{[B]}$ only.  

\begin{figure}[t]
\centering
\includegraphics[width=.48\textwidth]{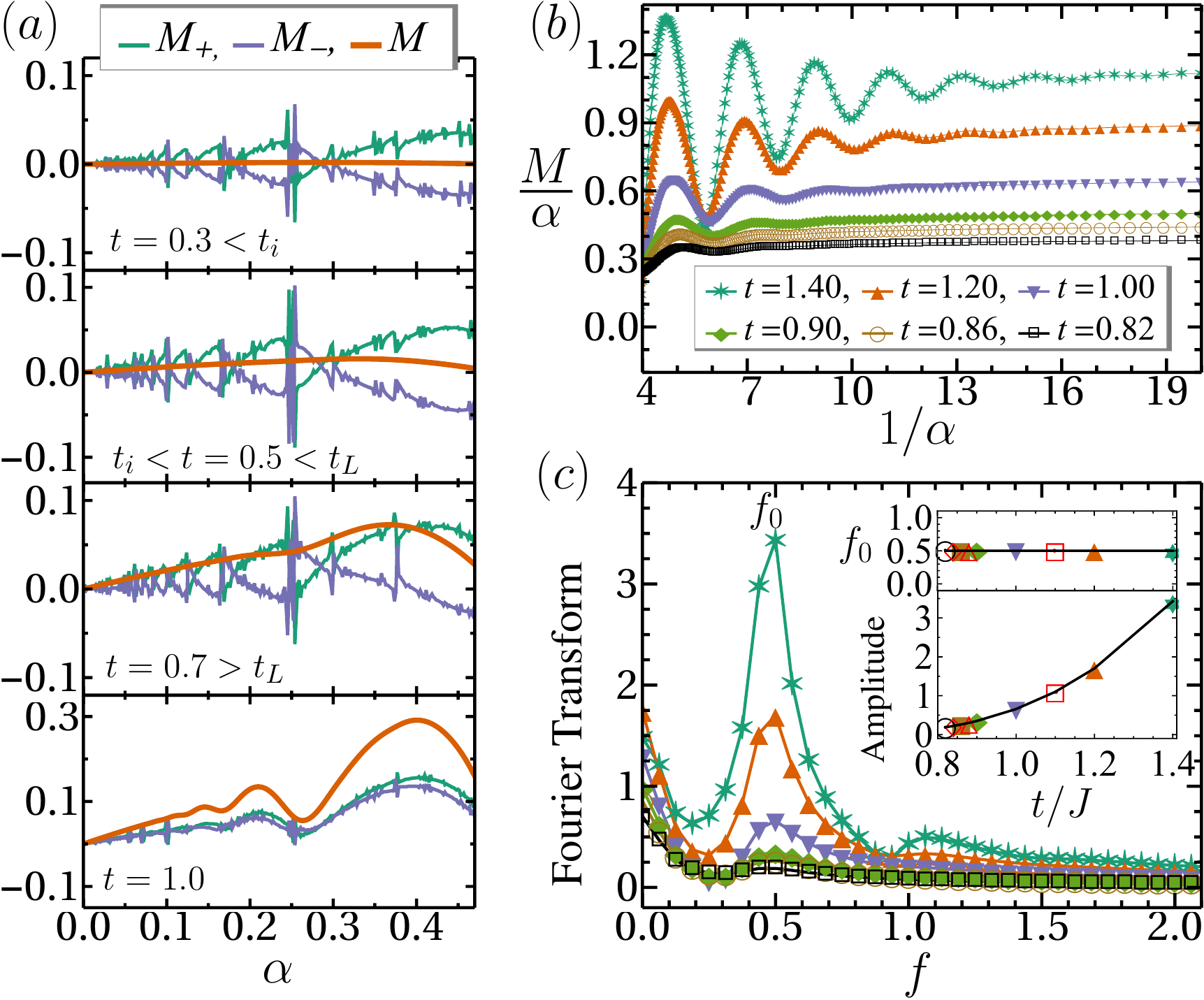}
\caption{
The dHvA oscillations  from Eq.~(\ref{eq:Hab-B}) in the Kondo insulating ground state on square lattice.  ($a$) Magnetization vs. $\alpha$, where $M_\pm$ are the contributions from the two charge quasiparticle bands, and $M=M_+ + M_-$.  ($b$) $M/\alpha$ vs. $1/\alpha$. ($c$) Fourier transform of $M/\alpha$,  with an inset showing the dominant frequency of oscillation, $f_0$, and its amplitude vs. $t/J$. The $f_0= 0.5$ is $t$ independent, and it corresponds to the area  of the half-BZ enclosed by the $|\gamma_\k|=0$ contour [see Fig.~\ref{fig:mf-parameters}($b$)]. 
}
\label{fig:mag-kondo-sqr}
\end{figure}

 In Fig.~\ref{fig:mag-kondo-sqr}($a$), we show the evolution of magnetization behaviour with $t$ on square lattice. For $t<t_i$, we see no quantum oscillations of $M$ with respect to $\alpha$, except an overall sinusoidal variation of  negligible magnitude. 
This is because the non-trivial oscillatory contribution to $M$ from $E_{\k-}$ states  ($M_-$) cancels that  ($M_+$)  from $E_{\k+}$. 
It is like two opposite cyclotron orbits from two oppositely curved dispersions cancelling each other. 
This cancellation gets weaker as $E_{\k-}$ starts inverting. But  only when $t$ is sufficiently above $t_L$, with $E_{\k-}$  nearly fully inverted, 
we begin to clearly see the oscillations of $M$ 
in the ground state of $\Hhat_{c}^{[B]}$. These oscillations 
 are weak in the Kondo singlet phase for $t\lesssim t_c$, but become pronounced when $t$ increases into the N\'eel phase, as Fig.~\ref{fig:mag-kondo-sqr}($b$) shows. The Fourier transform of $M/\alpha$  (with flat background subtracted) for $4<1/\alpha<20$ is presented in Fig.~\ref{fig:mag-kondo-sqr}($c$), where the dominant Fourier peaks for different $t$'s occur at the same frequency, $f_0 = 0.5$,  
while their amplitudes grow with $t$  [empirically, 
as $ (t-t_*)^2$ with $ t_* \approx 0.57 \gtrsim t_L$].

The semiclassical relation, $F=(2\pi/a)^2 f$, between the area $F$ of an extremal orbit perpendicular to magnetic field on a constant energy surface in $\k$-space and the frequency $f$ (in units of $h/e a^2$) of dHvA oscillations~\cite{Onsager}, implies that  the $f_0=0.5$ corresponds to the area of the half-BZ, which unmistakably points to the $|\gamma_\k|=0$ in Fig.~\ref{fig:mf-parameters}($b$) as its origin. From this, we infer that the dHvA oscillations in a KI measure the CPS of its charge quasiparticles~\footnote{ To resolve a possible $t$-dependent second frequency $\lesssim 0.5$ (corresponding to the second branch of CPS), one would need to compute on much larger systems.}.  
We think of the CPS as a generalization of the FS to the cases with gapped fermion quasiparticles. In the gapless Fermi systems, say metals, the CPS would be the Fermi surface. 

Similarly, we also get quantum oscillations of magnetization on simple cubic lattice, as shown in Fig.~\ref{fig:mag-kondo-cub}($a$). Its Fourier transform in Fig.~\ref{fig:mag-kondo-cub}($b$) gives the dominant frequency at $f_0=0.185$, which is independent of $t/J$, and corresponds precisely to the area enclosed by the blue-colored orbit shown in Fig.~\ref{fig:mag-kondo-cub}($c$). It is an extremal orbit on the $|\gamma_\k|=0$ branch of the CPS on $k_z=0$ plane. It is very clear that the dHvA oscillations measure the CPS, in corroboration of what we found on square lattice. 

\begin{figure}[t]
\centering
\includegraphics[width=.48\textwidth]{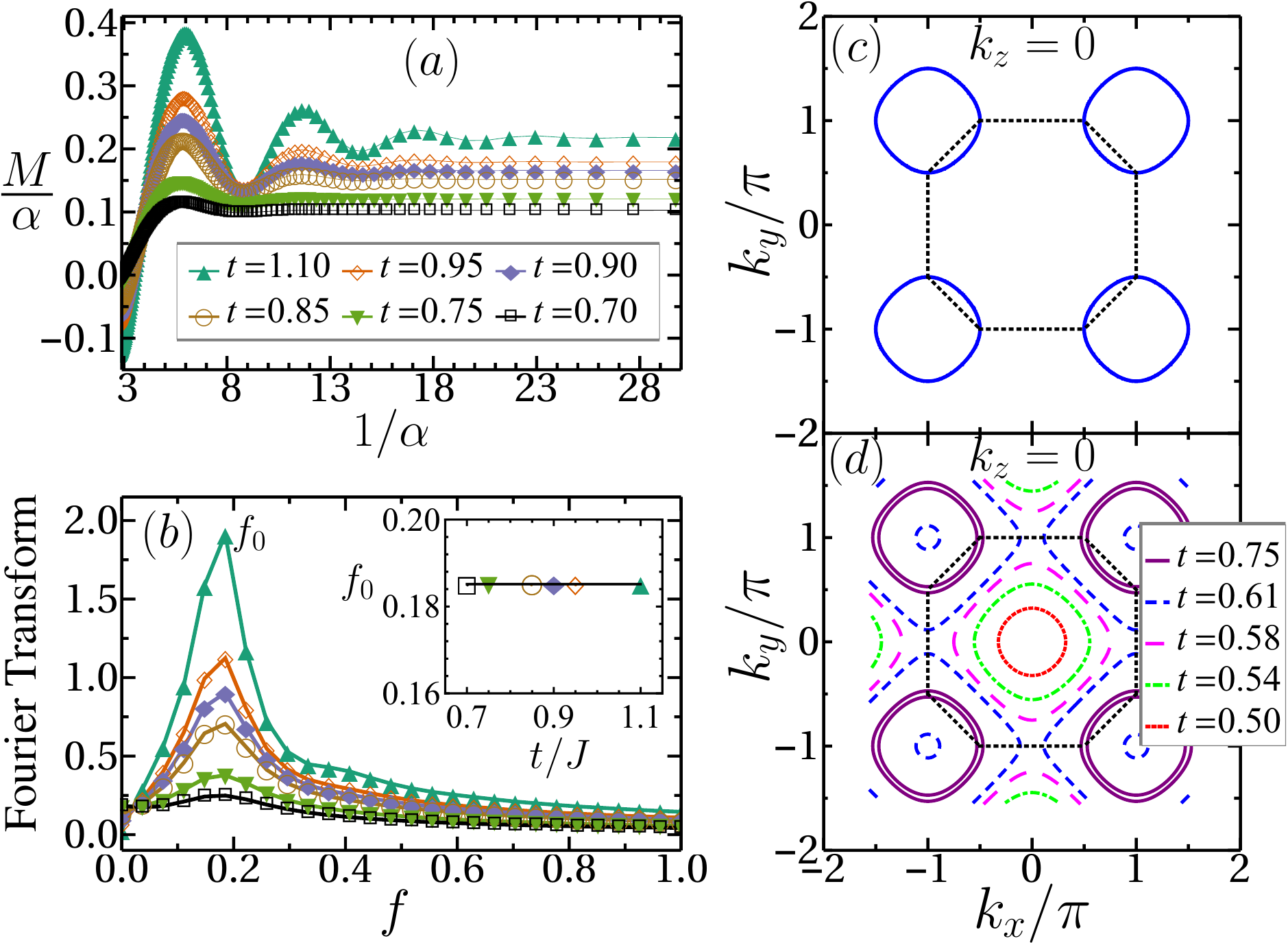}
\caption{($a$) ${M}/{\alpha}$ vs. ${1}/{\alpha}$ in the Kondo insulating ground state on simple cubic lattice. ($b$) Fourier transform of $M/\alpha$. The dominant frequency, $f_0 = 0.185$, is same as the area enclosed by the blue orbit in ($c$). It is a  $t$ independent extremal orbit on the $|\gamma_\k|=0$ CPS. 
($d$) The second branch of CPS. It tends to the first one with increasing $t$. The  dotted octagons 
in ($c$)-($d$) denote the boundary of the half-BZ on $k_z = 0$ plane.}
\label{fig:mag-kondo-cub}
\end{figure}

\begin{figure}[b]
\centering
\includegraphics[width=.48\textwidth]{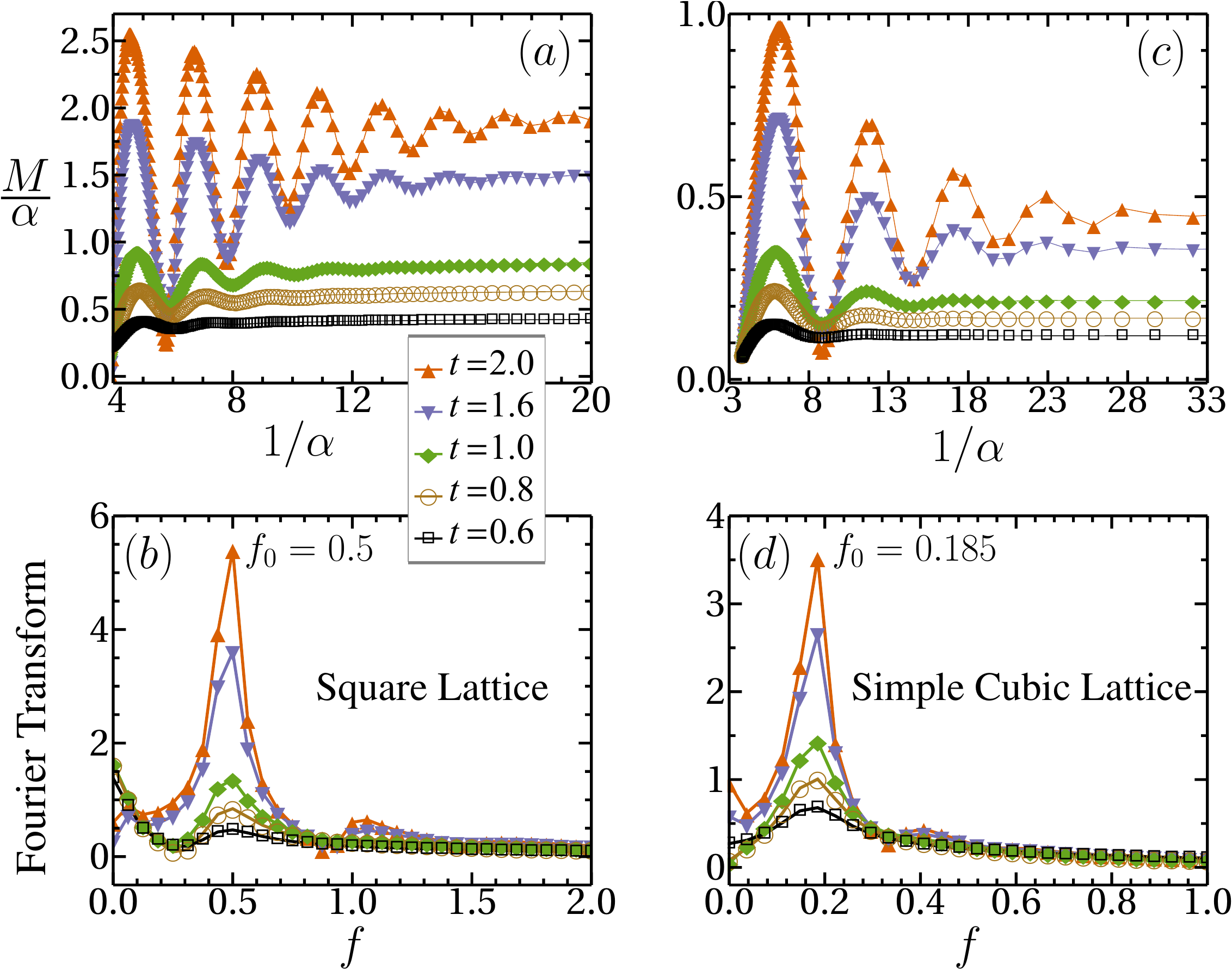}
\caption{The dHvA oscillations in the insulating N\'eel ground state of the Hubbard model (with $U=1$) 
at half-filling on ($a$) square and ($c$) simple cubic lattices. Their Fourier amplitudes (divided by $t$ for better visibility at smaller $t$'s) are plotted in ($b$) and ($d$). The dominant frequency, $f_0$, in the two cases here is same as that for the  corresponding KI's.}
\label{fig:hubbard}
\end{figure}


{\em Quantum oscillations in SDW insulators.---}
The above findings for the  KLM prompted us to also study dHvA oscillations in the Hubbard model, for which the present approach was invented~\cite{BK2008}. For small $t$, in units of the local repulsion $U$, the Mott insulating N\'eel ground state at half-filling on bipartite lattices is described here by the gapped, oppositely curved dispersions, $E_{\k\pm}$. We take $\rho_1=-1.338$ (quantum monte carlo value~\cite{Manousakis.RMP}) on square lattice, and $-1.194$ (spin-wave theory) on simple cubic lattice. Here, $E_{\k+}$ on square  (simple cubic) lattice starts inverting at $t_i = 0.016$  (0.007), and undergoes Lifshitz-like transition at {$t_L  \approx 2t_i $. For the same $\alpha=p/q$ as taken for KI's, the data in Fig.~\ref{fig:hubbard} shows clear oscillations for $t\gtrsim 0.5$, with $f_0= 0.5$ and $0.185$ coming from the $|\gamma_\k|=0$ CPS on square and simple cubic lattices. This calculation predicts the dHvA oscillations  to occur in spin-density wave (SDW) insulators, because the insulating state of the half-filled Hubbard model for such large values of $t$ describes the SDW insulators.


{\em Conclusion.---} 
To understand the quantum oscillations of magnetization in Kondo insulators, we have studied the spin-1/2 Kondo lattice model at half-filling on square and simple cubic lattices. The key finding of our study is that the dHvA oscillations in Kondo insulators  occur as a bulk phenomenon, which manifests itself through the inversion of a Hofstadter-quantized dispersion of the gapped charge quasiparticles whose chemical-potential surface these oscillations measure. We have found this through a  minimal effective dynamics,  in a certain canonical representation of electrons, that appropriately describes the Kondo insulating ground state, and reveals the inversion and Lifshitz-like transition for charge quasiparticles. This approach also gives the same oscillations in the N\'eel insulating ground state of the half-filled Hubbard model, with an amplitude that grows with hopping. It  clearly suggests that the spin-density wave insulators would also exhibit  quantum oscillations of magnetization. This needs to be investigated further, and will be discussed elsewhere. It would also be interesting to investigate the  quasiparticle band inversion, that we have found on bipartite lattices, on non-bipartite  Kondo lattices.


\begin{acknowledgments}
 We thank Sriram Shastry for useful suggestions. P.R. thanks CSIR (India) for financial support. B.K. acknowledges financial support under UPE-II and DST-PURSE programs of JNU,  and also acknowledges ICTP for an Associate visit during which some parts of this paper were written.
We also acknowledge the HPC cluster at IUAC, and DST-FIST funded HPC cluster at SPS, JNU.
\end{acknowledgments}

\bibliography{references.bib}

\begin{thebibliography}{35}%
\makeatletter
\providecommand \@ifxundefined [1]{%
 \@ifx{#1\undefined}
}%
\providecommand \@ifnum [1]{%
 \ifnum #1\expandafter \@firstoftwo
 \else \expandafter \@secondoftwo
 \fi
}%
\providecommand \@ifx [1]{%
 \ifx #1\expandafter \@firstoftwo
 \else \expandafter \@secondoftwo
 \fi
}%
\providecommand \natexlab [1]{#1}%
\providecommand \enquote  [1]{``#1''}%
\providecommand \bibnamefont  [1]{#1}%
\providecommand \bibfnamefont [1]{#1}%
\providecommand \citenamefont [1]{#1}%
\providecommand \href@noop [0]{\@secondoftwo}%
\providecommand \href [0]{\begingroup \@sanitize@url \@href}%
\providecommand \@href[1]{\@@startlink{#1}\@@href}%
\providecommand \@@href[1]{\endgroup#1\@@endlink}%
\providecommand \@sanitize@url [0]{\catcode `\\12\catcode `\$12\catcode
  `\&12\catcode `\#12\catcode `\^12\catcode `\_12\catcode `\%12\relax}%
\providecommand \@@startlink[1]{}%
\providecommand \@@endlink[0]{}%
\providecommand \url  [0]{\begingroup\@sanitize@url \@url }%
\providecommand \@url [1]{\endgroup\@href {#1}{\urlprefix }}%
\providecommand \urlprefix  [0]{URL }%
\providecommand \Eprint [0]{\href }%
\providecommand \doibase [0]{http://dx.doi.org/}%
\providecommand \selectlanguage [0]{\@gobble}%
\providecommand \bibinfo  [0]{\@secondoftwo}%
\providecommand \bibfield  [0]{\@secondoftwo}%
\providecommand \translation [1]{[#1]}%
\providecommand \BibitemOpen [0]{}%
\providecommand \bibitemStop [0]{}%
\providecommand \bibitemNoStop [0]{.\EOS\space}%
\providecommand \EOS [0]{\spacefactor3000\relax}%
\providecommand \BibitemShut  [1]{\csname bibitem#1\endcsname}%
\let\auto@bib@innerbib\@empty
\bibitem [{\citenamefont {Coleman}(2015)}]{Coleman.Book.2015}%
  \BibitemOpen
  \bibfield  {author} {\bibinfo {author} {\bibfnamefont {P.}~\bibnamefont
  {Coleman}},\ }\href@noop {} {\emph {\bibinfo {title} {Introduction to
  Many-Body Physics}}}\ (\bibinfo  {publisher} {Cambridge University Press,
  UK},\ \bibinfo {year} {2015})\BibitemShut {NoStop}%
\bibitem [{\citenamefont {Misra}(2008)}]{Misra.Book.2008}%
  \BibitemOpen
  \bibfield  {author} {\bibinfo {author} {\bibfnamefont {P.}~\bibnamefont
  {Misra}},\ }\href@noop {} {\emph {\bibinfo {title} {Heavy-Fermion Systems}}}\
  (\bibinfo  {publisher} {Elsevier, Amsterdam},\ \bibinfo {year}
  {2008})\BibitemShut {NoStop}%
\bibitem [{\citenamefont {Aeppli}\ and\ \citenamefont
  {Fisk}(1992)}]{Aeppli.Fisk.1992}%
  \BibitemOpen
  \bibfield  {author} {\bibinfo {author} {\bibfnamefont {G.}~\bibnamefont
  {Aeppli}}\ and\ \bibinfo {author} {\bibfnamefont {Z.}~\bibnamefont {Fisk}},\
  }\href@noop {} {\bibfield  {journal} {\bibinfo  {journal} {Comments Cond.
  Mat. Phys.}\ }\textbf {\bibinfo {volume} {16}},\ \bibinfo {pages} {155}
  (\bibinfo {year} {1992})}\BibitemShut {NoStop}%
\bibitem [{\citenamefont {Li}\ \emph {et~al.}(2014)\citenamefont {Li},
  \citenamefont {Xiang}, \citenamefont {Yu}, \citenamefont {Asaba},
  \citenamefont {Lawson}, \citenamefont {Cai}, \citenamefont {Tinsman},
  \citenamefont {Berkley}, \citenamefont {Wolgast}, \citenamefont {Eo},
  \citenamefont {Kim}, \citenamefont {Kurdak}, \citenamefont {Allen},
  \citenamefont {Sun}, \citenamefont {Chen}, \citenamefont {Wang},
  \citenamefont {Fisk},\ and\ \citenamefont {Li}}]{Li2014}%
  \BibitemOpen
  \bibfield  {author} {\bibinfo {author} {\bibfnamefont {G.}~\bibnamefont
  {Li}}, \bibinfo {author} {\bibfnamefont {Z.}~\bibnamefont {Xiang}}, \bibinfo
  {author} {\bibfnamefont {F.}~\bibnamefont {Yu}}, \bibinfo {author}
  {\bibfnamefont {T.}~\bibnamefont {Asaba}}, \bibinfo {author} {\bibfnamefont
  {B.}~\bibnamefont {Lawson}}, \bibinfo {author} {\bibfnamefont
  {P.}~\bibnamefont {Cai}}, \bibinfo {author} {\bibfnamefont {C.}~\bibnamefont
  {Tinsman}}, \bibinfo {author} {\bibfnamefont {A.}~\bibnamefont {Berkley}},
  \bibinfo {author} {\bibfnamefont {S.}~\bibnamefont {Wolgast}}, \bibinfo
  {author} {\bibfnamefont {Y.~S.}\ \bibnamefont {Eo}}, \bibinfo {author}
  {\bibfnamefont {D.-J.}\ \bibnamefont {Kim}}, \bibinfo {author} {\bibfnamefont
  {C.}~\bibnamefont {Kurdak}}, \bibinfo {author} {\bibfnamefont {J.~W.}\
  \bibnamefont {Allen}}, \bibinfo {author} {\bibfnamefont {K.}~\bibnamefont
  {Sun}}, \bibinfo {author} {\bibfnamefont {X.~H.}\ \bibnamefont {Chen}},
  \bibinfo {author} {\bibfnamefont {Y.~Y.}\ \bibnamefont {Wang}}, \bibinfo
  {author} {\bibfnamefont {Z.}~\bibnamefont {Fisk}}, \ and\ \bibinfo {author}
  {\bibfnamefont {L.}~\bibnamefont {Li}},\ }\href {\doibase
  10.1126/science.1250366} {\bibfield  {journal} {\bibinfo  {journal}
  {Science}\ }\textbf {\bibinfo {volume} {346}},\ \bibinfo {pages} {1208}
  (\bibinfo {year} {2014})}\BibitemShut {NoStop}%
\bibitem [{\citenamefont {Tan}\ \emph {et~al.}(2015)\citenamefont {Tan},
  \citenamefont {Hsu}, \citenamefont {Zeng}, \citenamefont {Hatnean},
  \citenamefont {Harrison}, \citenamefont {Zhu}, \citenamefont {Hartstein},
  \citenamefont {Kiourlappou}, \citenamefont {Srivastava}, \citenamefont
  {Johannes}, \citenamefont {Murphy}, \citenamefont {Park}, \citenamefont
  {Balicas}, \citenamefont {Lonzarich}, \citenamefont {Balakrishnan},\ and\
  \citenamefont {Sebastian}}]{Tan2015}%
  \BibitemOpen
  \bibfield  {author} {\bibinfo {author} {\bibfnamefont {B.~S.}\ \bibnamefont
  {Tan}}, \bibinfo {author} {\bibfnamefont {Y.-T.}\ \bibnamefont {Hsu}},
  \bibinfo {author} {\bibfnamefont {B.}~\bibnamefont {Zeng}}, \bibinfo {author}
  {\bibfnamefont {M.~C.}\ \bibnamefont {Hatnean}}, \bibinfo {author}
  {\bibfnamefont {N.}~\bibnamefont {Harrison}}, \bibinfo {author}
  {\bibfnamefont {Z.}~\bibnamefont {Zhu}}, \bibinfo {author} {\bibfnamefont
  {M.}~\bibnamefont {Hartstein}}, \bibinfo {author} {\bibfnamefont
  {M.}~\bibnamefont {Kiourlappou}}, \bibinfo {author} {\bibfnamefont
  {A.}~\bibnamefont {Srivastava}}, \bibinfo {author} {\bibfnamefont {M.~D.}\
  \bibnamefont {Johannes}}, \bibinfo {author} {\bibfnamefont {T.~P.}\
  \bibnamefont {Murphy}}, \bibinfo {author} {\bibfnamefont {J.-H.}\
  \bibnamefont {Park}}, \bibinfo {author} {\bibfnamefont {L.}~\bibnamefont
  {Balicas}}, \bibinfo {author} {\bibfnamefont {G.~G.}\ \bibnamefont
  {Lonzarich}}, \bibinfo {author} {\bibfnamefont {G.}~\bibnamefont
  {Balakrishnan}}, \ and\ \bibinfo {author} {\bibfnamefont {S.~E.}\
  \bibnamefont {Sebastian}},\ }\href {\doibase 10.1126/science.aaa7974}
  {\bibfield  {journal} {\bibinfo  {journal} {Science}\ }\textbf {\bibinfo
  {volume} {349}},\ \bibinfo {pages} {287} (\bibinfo {year}
  {2015})}\BibitemShut {NoStop}%
\bibitem [{\citenamefont {Ashcroft}\ and\ \citenamefont
  {Mermin}(1976)}]{Ashcroft.Mermin}%
  \BibitemOpen
  \bibfield  {author} {\bibinfo {author} {\bibfnamefont {N.~W.}\ \bibnamefont
  {Ashcroft}}\ and\ \bibinfo {author} {\bibfnamefont {N.~D.}\ \bibnamefont
  {Mermin}},\ }\href@noop {} {\emph {\bibinfo {title} {Solid State Physics}}}\
  (\bibinfo  {publisher} {Saunders College Publishing, USA},\ \bibinfo {year}
  {1976})\BibitemShut {NoStop}%
\bibitem [{\citenamefont {Abrikosov}(1988)}]{Abrikosov1988}%
  \BibitemOpen
  \bibfield  {author} {\bibinfo {author} {\bibfnamefont {A.~A.}\ \bibnamefont
  {Abrikosov}},\ }\href@noop {} {\emph {\bibinfo {title} {Fundamentals of the
  Theory of Metals}}}\ (\bibinfo  {publisher} {North-Holland, Amsterdam},\
  \bibinfo {year} {1988})\BibitemShut {NoStop}%
\bibitem [{\citenamefont {Onsager}(1952)}]{Onsager}%
  \BibitemOpen
  \bibfield  {author} {\bibinfo {author} {\bibfnamefont {L.}~\bibnamefont
  {Onsager}},\ }\href {\doibase 10.1080/14786440908521019} {\bibfield
  {journal} {\bibinfo  {journal} {Phil. Mag.}\ }\textbf {\bibinfo {volume}
  {43}},\ \bibinfo {pages} {1006} (\bibinfo {year} {1952})}\BibitemShut
  {NoStop}%
\bibitem [{\citenamefont {Kishigi}\ and\ \citenamefont
  {Hasegawa}(2014)}]{Kishigi2014}%
  \BibitemOpen
  \bibfield  {author} {\bibinfo {author} {\bibfnamefont {K.}~\bibnamefont
  {Kishigi}}\ and\ \bibinfo {author} {\bibfnamefont {Y.}~\bibnamefont
  {Hasegawa}},\ }\href {\doibase 10.1103/PhysRevB.90.085427} {\bibfield
  {journal} {\bibinfo  {journal} {Phys. Rev. B}\ }\textbf {\bibinfo {volume}
  {90}},\ \bibinfo {pages} {085427} (\bibinfo {year} {2014})}\BibitemShut
  {NoStop}%
\bibitem [{\citenamefont {Knolle}\ and\ \citenamefont
  {Cooper}(2015)}]{Knolle2015}%
  \BibitemOpen
  \bibfield  {author} {\bibinfo {author} {\bibfnamefont {J.}~\bibnamefont
  {Knolle}}\ and\ \bibinfo {author} {\bibfnamefont {N.~R.}\ \bibnamefont
  {Cooper}},\ }\href {\doibase 10.1103/PhysRevLett.115.146401} {\bibfield
  {journal} {\bibinfo  {journal} {Phys. Rev. Lett.}\ }\textbf {\bibinfo
  {volume} {115}},\ \bibinfo {pages} {146401} (\bibinfo {year}
  {2015})}\BibitemShut {NoStop}%
\bibitem [{\citenamefont {Zhang}\ \emph {et~al.}(2016)\citenamefont {Zhang},
  \citenamefont {Song},\ and\ \citenamefont {Wang}}]{Zhang2016}%
  \BibitemOpen
  \bibfield  {author} {\bibinfo {author} {\bibfnamefont {L.}~\bibnamefont
  {Zhang}}, \bibinfo {author} {\bibfnamefont {X.-Y.}\ \bibnamefont {Song}}, \
  and\ \bibinfo {author} {\bibfnamefont {F.}~\bibnamefont {Wang}},\ }\href
  {\doibase 10.1103/PhysRevLett.116.046404} {\bibfield  {journal} {\bibinfo
  {journal} {Phys. Rev. Lett.}\ }\textbf {\bibinfo {volume} {116}},\ \bibinfo
  {pages} {046404} (\bibinfo {year} {2016})}\BibitemShut {NoStop}%
\bibitem [{\citenamefont {Baskaran}(2015)}]{Baskaran2015}%
  \BibitemOpen
  \bibfield  {author} {\bibinfo {author} {\bibfnamefont {G.}~\bibnamefont
  {Baskaran}},\ }\href@noop {} {\bibfield  {journal} {\bibinfo  {journal}
  {arXiv:1507.03477}\ } (\bibinfo {year} {2015})}\BibitemShut {NoStop}%
\bibitem [{\citenamefont {Erten}\ \emph {et~al.}(2016)\citenamefont {Erten},
  \citenamefont {Ghaemi},\ and\ \citenamefont {Coleman}}]{Erten2016}%
  \BibitemOpen
  \bibfield  {author} {\bibinfo {author} {\bibfnamefont {O.}~\bibnamefont
  {Erten}}, \bibinfo {author} {\bibfnamefont {P.}~\bibnamefont {Ghaemi}}, \
  and\ \bibinfo {author} {\bibfnamefont {P.}~\bibnamefont {Coleman}},\ }\href
  {\doibase 10.1103/PhysRevLett.116.046403} {\bibfield  {journal} {\bibinfo
  {journal} {Phys. Rev. Lett.}\ }\textbf {\bibinfo {volume} {116}},\ \bibinfo
  {pages} {046403} (\bibinfo {year} {2016})}\BibitemShut {NoStop}%
\bibitem [{\citenamefont {Pal}\ \emph {et~al.}(2016)\citenamefont {Pal},
  \citenamefont {Pi\'echon}, \citenamefont {Fuchs}, \citenamefont {Goerbig},\
  and\ \citenamefont {Montambaux}}]{Pal2016}%
  \BibitemOpen
  \bibfield  {author} {\bibinfo {author} {\bibfnamefont {H.~K.}\ \bibnamefont
  {Pal}}, \bibinfo {author} {\bibfnamefont {F.}~\bibnamefont {Pi\'echon}},
  \bibinfo {author} {\bibfnamefont {J.-N.}\ \bibnamefont {Fuchs}}, \bibinfo
  {author} {\bibfnamefont {M.}~\bibnamefont {Goerbig}}, \ and\ \bibinfo
  {author} {\bibfnamefont {G.}~\bibnamefont {Montambaux}},\ }\href {\doibase
  10.1103/PhysRevB.94.125140} {\bibfield  {journal} {\bibinfo  {journal} {Phys.
  Rev. B}\ }\textbf {\bibinfo {volume} {94}},\ \bibinfo {pages} {125140}
  (\bibinfo {year} {2016})}\BibitemShut {NoStop}%
\bibitem [{\citenamefont {Dzero}\ \emph {et~al.}(2016)\citenamefont {Dzero},
  \citenamefont {Xia}, \citenamefont {Galitski},\ and\ \citenamefont
  {Coleman}}]{Dzero2016}%
  \BibitemOpen
  \bibfield  {author} {\bibinfo {author} {\bibfnamefont {M.}~\bibnamefont
  {Dzero}}, \bibinfo {author} {\bibfnamefont {J.}~\bibnamefont {Xia}}, \bibinfo
  {author} {\bibfnamefont {V.}~\bibnamefont {Galitski}}, \ and\ \bibinfo
  {author} {\bibfnamefont {P.}~\bibnamefont {Coleman}},\ }\href {\doibase
  10.1146/annurev-conmatphys-031214-014749} {\bibfield  {journal} {\bibinfo
  {journal} {Annu. Rev. Condens. Matter Phys.}\ }\textbf {\bibinfo {volume}
  {7}},\ \bibinfo {pages} {249} (\bibinfo {year} {2016})}\BibitemShut {NoStop}%
\bibitem [{Note1()}]{Note1}%
  \BibitemOpen
  \bibinfo {note} {The Fermi sea of non-interacting electrons on bipartite
  lattice, which is a conducting state, consists of four such independent
  gapless Majorana Fermi seas.}\BibitemShut {Stop}%
\bibitem [{\citenamefont {Xu}\ \emph {et~al.}(2016)\citenamefont {Xu},
  \citenamefont {Cui}, \citenamefont {Dong}, \citenamefont {Zhao},
  \citenamefont {Wu}, \citenamefont {Chen}, \citenamefont {Sun}, \citenamefont
  {Yao},\ and\ \citenamefont {Li}}]{Xu2016}%
  \BibitemOpen
  \bibfield  {author} {\bibinfo {author} {\bibfnamefont {Y.}~\bibnamefont
  {Xu}}, \bibinfo {author} {\bibfnamefont {S.}~\bibnamefont {Cui}}, \bibinfo
  {author} {\bibfnamefont {J.~K.}\ \bibnamefont {Dong}}, \bibinfo {author}
  {\bibfnamefont {D.}~\bibnamefont {Zhao}}, \bibinfo {author} {\bibfnamefont
  {T.}~\bibnamefont {Wu}}, \bibinfo {author} {\bibfnamefont {X.~H.}\
  \bibnamefont {Chen}}, \bibinfo {author} {\bibfnamefont {K.}~\bibnamefont
  {Sun}}, \bibinfo {author} {\bibfnamefont {H.}~\bibnamefont {Yao}}, \ and\
  \bibinfo {author} {\bibfnamefont {S.~Y.}\ \bibnamefont {Li}},\ }\href
  {\doibase 10.1103/PhysRevLett.116.246403} {\bibfield  {journal} {\bibinfo
  {journal} {Phys. Rev. Lett.}\ }\textbf {\bibinfo {volume} {116}},\ \bibinfo
  {pages} {246403} (\bibinfo {year} {2016})}\BibitemShut {NoStop}%
\bibitem [{\citenamefont {Kumar}(2008)}]{BK2008}%
  \BibitemOpen
  \bibfield  {author} {\bibinfo {author} {\bibfnamefont {B.}~\bibnamefont
  {Kumar}},\ }\href {\doibase 10.1103/PhysRevB.77.205115} {\bibfield  {journal}
  {\bibinfo  {journal} {Phys. Rev. B}\ }\textbf {\bibinfo {volume} {77}},\
  \bibinfo {pages} {205115} (\bibinfo {year} {2008})}\BibitemShut {NoStop}%
\bibitem [{\citenamefont {Kumar}(2009)}]{BK2009}%
  \BibitemOpen
  \bibfield  {author} {\bibinfo {author} {\bibfnamefont {B.}~\bibnamefont
  {Kumar}},\ }\href {\doibase 10.1103/PhysRevB.79.155121} {\bibfield  {journal}
  {\bibinfo  {journal} {Phys. Rev. B}\ }\textbf {\bibinfo {volume} {79}},\
  \bibinfo {pages} {155121} (\bibinfo {year} {2009})}\BibitemShut {NoStop}%
\bibitem [{\citenamefont {Kumar}(2013)}]{BK2013}%
  \BibitemOpen
  \bibfield  {author} {\bibinfo {author} {\bibfnamefont {B.}~\bibnamefont
  {Kumar}},\ }\href {\doibase 10.1103/PhysRevB.87.195105} {\bibfield  {journal}
  {\bibinfo  {journal} {Phys. Rev. B}\ }\textbf {\bibinfo {volume} {87}},\
  \bibinfo {pages} {195105} (\bibinfo {year} {2013})}\BibitemShut {NoStop}%
\bibitem [{Note2()}]{Note2}%
  \BibitemOpen
  \bibinfo {note} {The term corresponding to $t$ in $\protect \mathaccentV
  {hat}05E{H}_0$ is the so-called scalar Majorana Fermi sea of Ref.~\protect
  \rev@citealp {Baskaran2015}. But here it occurs with an additional term due
  to $J$ that opens the charge gap.}\BibitemShut {Stop}%
\bibitem [{Note3()}]{Note3}%
  \BibitemOpen
  \bibinfo {note} {The $\protect \mathaccentV {hat}05E{H}_{s}$ resembles the
  Kondo necklace model~\cite {Doniach-necklace}. Here, it describes the
  magnetic properties of KI.}\BibitemShut {Stop}%
\bibitem [{\citenamefont {Sachdev}\ and\ \citenamefont
  {Bhatt}(1990)}]{SachdevBhatt}%
  \BibitemOpen
  \bibfield  {author} {\bibinfo {author} {\bibfnamefont {S.}~\bibnamefont
  {Sachdev}}\ and\ \bibinfo {author} {\bibfnamefont {R.~N.}\ \bibnamefont
  {Bhatt}},\ }\href {\doibase 10.1103/PhysRevB.41.9323} {\bibfield  {journal}
  {\bibinfo  {journal} {Phys. Rev. B}\ }\textbf {\bibinfo {volume} {41}},\
  \bibinfo {pages} {9323} (\bibinfo {year} {1990})}\BibitemShut {NoStop}%
\bibitem [{\citenamefont {Kumar}(2010)}]{BK2010}%
  \BibitemOpen
  \bibfield  {author} {\bibinfo {author} {\bibfnamefont {B.}~\bibnamefont
  {Kumar}},\ }\href {\doibase 10.1103/PhysRevB.82.054404} {\bibfield  {journal}
  {\bibinfo  {journal} {Phys. Rev. B}\ }\textbf {\bibinfo {volume} {82}},\
  \bibinfo {pages} {054404} (\bibinfo {year} {2010})}\BibitemShut {NoStop}%
\bibitem [{\citenamefont {Assaad}(1999)}]{Assaad1999}%
  \BibitemOpen
  \bibfield  {author} {\bibinfo {author} {\bibfnamefont {F.~F.}\ \bibnamefont
  {Assaad}},\ }\href {\doibase 10.1103/PhysRevLett.83.796} {\bibfield
  {journal} {\bibinfo  {journal} {Phys. Rev. Lett.}\ }\textbf {\bibinfo
  {volume} {83}},\ \bibinfo {pages} {796} (\bibinfo {year} {1999})}\BibitemShut
  {NoStop}%
\bibitem [{\citenamefont {Shi}\ \emph {et~al.}(1995)\citenamefont {Shi},
  \citenamefont {Singh}, \citenamefont {Gelfand},\ and\ \citenamefont
  {Wang}}]{Shi.Singh.1995}%
  \BibitemOpen
  \bibfield  {author} {\bibinfo {author} {\bibfnamefont {Z.-P.}\ \bibnamefont
  {Shi}}, \bibinfo {author} {\bibfnamefont {R.~R.~P.}\ \bibnamefont {Singh}},
  \bibinfo {author} {\bibfnamefont {M.~P.}\ \bibnamefont {Gelfand}}, \ and\
  \bibinfo {author} {\bibfnamefont {Z.}~\bibnamefont {Wang}},\ }\href {\doibase
  10.1103/PhysRevB.51.15630} {\bibfield  {journal} {\bibinfo  {journal} {Phys.
  Rev. B}\ }\textbf {\bibinfo {volume} {51}},\ \bibinfo {pages} {15630}
  (\bibinfo {year} {1995})}\BibitemShut {NoStop}%
\bibitem [{\citenamefont {Wang}\ \emph {et~al.}(1994)\citenamefont {Wang},
  \citenamefont {Li},\ and\ \citenamefont {Lee}}]{Wang.Li.Lee.1994}%
  \BibitemOpen
  \bibfield  {author} {\bibinfo {author} {\bibfnamefont {Z.}~\bibnamefont
  {Wang}}, \bibinfo {author} {\bibfnamefont {X.-P.}\ \bibnamefont {Li}}, \ and\
  \bibinfo {author} {\bibfnamefont {D.-H.}\ \bibnamefont {Lee}},\ }\href@noop
  {} {\bibfield  {journal} {\bibinfo  {journal} {Physica B}\ }\textbf {\bibinfo
  {volume} {199 \& 200}},\ \bibinfo {pages} {463} (\bibinfo {year}
  {1994})}\BibitemShut {NoStop}%
\bibitem [{Note4()}]{Note4}%
  \BibitemOpen
  \bibinfo {note} {No such inversion occurs for the triplon dispersion,
  $\varepsilon _{\protect \bf k}$.}\BibitemShut {Stop}%
\bibitem [{\citenamefont {Trebst}\ \emph {et~al.}(2006)\citenamefont {Trebst},
  \citenamefont {Monien}, \citenamefont {Grzesik},\ and\ \citenamefont
  {Sigrist}}]{Trebst2006}%
  \BibitemOpen
  \bibfield  {author} {\bibinfo {author} {\bibfnamefont {S.}~\bibnamefont
  {Trebst}}, \bibinfo {author} {\bibfnamefont {H.}~\bibnamefont {Monien}},
  \bibinfo {author} {\bibfnamefont {A.}~\bibnamefont {Grzesik}}, \ and\
  \bibinfo {author} {\bibfnamefont {M.}~\bibnamefont {Sigrist}},\ }\href
  {\doibase 10.1103/PhysRevB.73.165101} {\bibfield  {journal} {\bibinfo
  {journal} {Phys. Rev. B}\ }\textbf {\bibinfo {volume} {73}},\ \bibinfo
  {pages} {165101} (\bibinfo {year} {2006})}\BibitemShut {NoStop}%
\bibitem [{\citenamefont {Lifshitz}(1960)}]{Lifshitz}%
  \BibitemOpen
  \bibfield  {author} {\bibinfo {author} {\bibfnamefont {I.~M.}\ \bibnamefont
  {Lifshitz}},\ }\href@noop {} {\bibfield  {journal} {\bibinfo  {journal} {Sov.
  Phys. JETP}\ }\textbf {\bibinfo {volume} {11}},\ \bibinfo {pages} {1130}
  (\bibinfo {year} {1960})}\BibitemShut {NoStop}%
\bibitem [{\citenamefont {Hofstadter}(1976)}]{Hofstadter1976}%
  \BibitemOpen
  \bibfield  {author} {\bibinfo {author} {\bibfnamefont {D.~R.}\ \bibnamefont
  {Hofstadter}},\ }\href {\doibase 10.1103/PhysRevB.14.2239} {\bibfield
  {journal} {\bibinfo  {journal} {Phys. Rev. B}\ }\textbf {\bibinfo {volume}
  {14}},\ \bibinfo {pages} {2239} (\bibinfo {year} {1976})}\BibitemShut
  {NoStop}%
\bibitem [{Note5()}]{Note5}%
  \BibitemOpen
  \bibinfo {note} {Unlike the basic Hofstadter model, in our $H_c^{[B]}$ (that
  has hopping and pairing), the Landau bands are somewhat dispersive with
  respect to $k_x$, even for large $q$. Hence, in our calculations, we have
  also taken upto 288 $k_x$-points.}\BibitemShut {Stop}%
\bibitem [{Note6()}]{Note6}%
  \BibitemOpen
  \bibinfo {note} {To resolve a possible $t$-dependent second frequency
  $\lesssim 0.5$ (corresponding to the second branch of CPS), one would need to
  compute on much larger systems.}\BibitemShut {Stop}%
\bibitem [{\citenamefont {Manousakis}(1991)}]{Manousakis.RMP}%
  \BibitemOpen
  \bibfield  {author} {\bibinfo {author} {\bibfnamefont {E.}~\bibnamefont
  {Manousakis}},\ }\href {\doibase 10.1103/RevModPhys.63.1} {\bibfield
  {journal} {\bibinfo  {journal} {Rev. Mod. Phys.}\ }\textbf {\bibinfo {volume}
  {63}},\ \bibinfo {pages} {1} (\bibinfo {year} {1991})}\BibitemShut {NoStop}%
\bibitem [{\citenamefont {Doniach}(1977)}]{Doniach-necklace}%
  \BibitemOpen
  \bibfield  {author} {\bibinfo {author} {\bibfnamefont {S.}~\bibnamefont
  {Doniach}},\ }\href@noop {} {\bibfield  {journal} {\bibinfo  {journal}
  {Physica B}\ }\textbf {\bibinfo {volume} {91}},\ \bibinfo {pages} {231}
  (\bibinfo {year} {1977})}\BibitemShut {NoStop}%
\end{thebibliography}%

\end{document}